%% file: MF479rv.tex
\documentclass[useAMS,usenatbib]{mn2e}

\usepackage{graphicx}
\usepackage{amssymb}
\usepackage{latexsym}
\usepackage{amsmath}
\usepackage{astro_bib_macro}
\usepackage{natbib}

\newcommand{\du}{\mathrm{d}}

\newcommand{\unldots}{..}
\newcommand{\deuxvdots}{:}

\begin{document}

\title[Comparison of Fourier and model-based estimators in single mode
  multiaxial interferometry]{Comparison of Fourier and model-based estimators in single mode
  multiaxial interferometry}
\author[E. Tatulli and J.-B. LeBouquin]{E. Tatulli$^{1}$\thanks{E-mail:
lastname@obs.ujf-grenoble.fr} and J.-B. LeBouquin$^{1}$\\
$^{1}$Laboratoire d'Astrophysique, Observatoire de Grenoble, 38041 Grenoble cedex France}

\date{to be inserted later}

\pagerange{\pageref{firstpage}--\pageref{lastpage}} \pubyear{2005}

\maketitle

\label{firstpage}

\begin{abstract}
There are several solutions to code the signal arising from
  optical long baseline multi-aperture interferometers. In this paper,
  we focus on the {\bf non homothetic spatial coding scheme} (multiaxial) with  the fringe pattern coded along one dimension on one detector
  (all-in-one). After describing the physical principles governing 
  single mode interferometers using that sort of recombination scheme, 
  we analyze 
  two different existing methods that measure the source visibility.
  The first technique, so-called Fourier estimator,
  consists in integrating the high frequency peak
  of the {\bf power} spectral density of the interferogram. The second method,
  so-called model-based estimator, 
  has been specifically developed for the AMBER instrument of the VLTI
  and deals with 
  directly modelling the interferogram recorded on the
  detector. Performances of both estimators are computed in terms of
  Signal to Noise Ratio (SNR) of the visibility, assuming that the
  interferograms are perturbed by photon and detector noises. 
  Theoretical expressions of the visibility SNR are provided, validated
  through numerical computations and then compared. We show that the
  model-based estimator offers up to $5$ times better
  performances than the Fourier one.    
 \end{abstract}


\section{Introduction}
 The next challenge of long baseline optical interferometry 
 is to commonly perform direct imaging
 of the observed source, the analogous way it is 
 done in radio-interferometry \citep{hogbom_1} 
or in infrared aperture masking \citep{tuthill_etal_1}. 
 After the first promising results obtained with {\bf COAST \citep{baldwin_1, young_1}, NPOI
 \citep{hummel_1} and IOTA \citep{monnier_1}}, such technique should soon move
 one step forward with the operating of the 
 AMBER instrument \citep{petrov_etal_1},
 the three beam recombiner of the VLTI. From the beginning of 2005,
 AMBER will indeed take full benefit of the
 unique combination of the great sensitivity of large aperture
 telescopes and the spatial frequency coverage provided by the VLTI, 
 even though it will require {\bf multiple nights of observing} to be able to restore consistent
 images \citep{thiebaut_1, tatulli_etal_2}. Then, in less than a
 decade, huge improvements are
 contemplated to be accomplished with second
 generation instruments of the VLTI that will enable 
snapshot imaging by using $4$, $6$ or even  $8$ telescopes
  simultaneously {\bf (e.g. \citet{malbet_etal_1})}.

One critical point in the design of future interferometric imaging
instruments is the choice of the beam recombination scheme, which can
become particularly complex, 
especially when dealing with multi apertures ($N_{tel} \ge 3$) interferometers. Following the solution that 
has been chosen for the AMBER instrument, we investigate 
the properties of single mode non-homothetic {\bf spatial coding} scheme (from now on "multiaxial") with all
the fringes pattern in the same spatial dimension on the same detector (from
now on "all-in-one"). In other words, interferograms are obtained by
mixing all together the input beams arising
from the different telescopes, thanks to output
pupils arranged along one single dimension (see Fig. \ref{fig_dessin_interf}).  We analyze the ways to estimate the source
visibility from such interferograms. 

Indeed, single mode multiaxial all-in-one recombination appears particularly
well suited in the framework of interferometric imaging. 
First it is the simplest and 
most compact way to recover informations arising from all the baselines. 
Moreover
it provides a better transmission than Michelson
recombination schemes (i.e temporal coding) since it makes use of less mirrors
and beam splitter for the same given number of input pupils. And the number of
pixels required to code the signal is also smaller, which drives to
higher limiting magnitudes \citep{jblebou_etal_2}. 
Furthermore, the remarkable
spatial filtering properties of single mode waveguides allow to
change the phase corrugations of the incoming turbulent wavefront into
intensity fluctuations at the output of the fibers. In other words,
only one fraction of the source flux, so-called coupling coefficient which
depends on the Strehl ratio {\bf of the pupil apodized by the fiber single mode} \citep{foresto_1}, remains in the interferogram.      
But the very advantageous counterpart is that the
shape of the interferogram is entirely deterministic, {\bf that is the form of the peaks in the Fourier plane is fixed, and the fringe pattern fully determined by two free parameters, its amplitude and its phase.}

Before the advent of single mode interferometers and in 
order to overcome the problem of
the turbulence, 
\citet{roddier_lena_1} proposed to estimate the
visibility in the Fourier plane from the
integration of the high frequency peak of the long exposure {\bf power} spectral density of the interferogram. Then \citet{foresto_2}, in a natural
way, used the same estimator to compute the visibility arising from the
FLUOR experiment, the first interferometer making use of
single mode waveguides. In multiaxial coding, we can furthermore take
advantage of the deterministic nature of the interferogram shape {\bf to perform model fitting techniques}.
It was though only recently, namely for the AMBER
instrument, that this property was used  
to estimate the visibility, by fitting the fringe pattern {\bf-- its phase and its amplitude --} in
the detector plane \citep{millour_etal_1}.

In this paper, we recall in Section \ref{sec_multi} 
the general formalism of multiaxial all-in-one 
recombination, that is the equation governing 
the interferogram, as well as the two techniques currently used to estimate the
visibility. As mentioned above, the first technique, so-called Fourier
estimator, integrates the high frequency peak of the
{\bf power} spectral density of the interferogram whereas the second one,
so-called model-based estimator, directly models 
the interferogram in the detector plane. 
For both estimators, theoretical
expressions of Signal to Noise Ratio (SNR) of the visibility are provided.
In Section \ref{sec_valid}, those expressions are validated thanks to
numerical simulations of noisy interferograms.
Then in Section \ref{sec_compare} both estimators are compared, from a
formal point of view and in terms of relative 
performances. The influence of 
instrumental parameters is investigated as well, with a special emphasis
regarding the choice of the width of the reading window of the detector.  {\bf The presence of atmospheric piston which blurs the fringes has not been taken into account in this analysis. Indeed, regardless the chosen estimator, its effect results in an attenuation of the squared visibility \citep{colavita_1}, and the sensitivity of both estimators to this point is the same.}   

This paper is the first part of our study on 
mutiaxial all-in-one recombination.  
In a second paper \citep{jblebou_etal_1}, we
analyze in which way such recombination scheme allows to
optimize the visibility SNR thanks to specific geometric configuration
of the output pupils.

\section{General formalism} \label{sec_multi}
\begin{figure}
    \includegraphics[width=8cm,height=5cm]{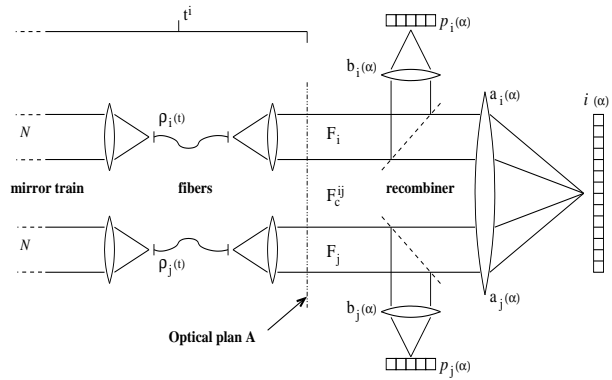}
    \caption{Sketch of a
        multiaxial all-in-one single mode interferometer. $t^i$
        denotes the total "static" transmission from the $i^{th}$
        telescope aperture to the optical plane A (i.e. mirrors,
        delay line, transmission of the $i^{th}$ fiber, ...), whereas $\rho^i$
        takes into account the "dynamical" transmission, that is the
        coupling coefficient of the $i^{th}$ fiber.     
    $i(\alpha)$ is the interferogram. $p^i(\alpha)$ and $p^j(\alpha)$ are
the photometric channels. The photometric fluxes $F^i$,
        $F^j$ and the coherent flux $F_c^{ij}$ are defined in the
        optical plane A. 
$a^i(\alpha)$ and $b^i(\alpha)$ are respectively the detected beams in
the interferometric and photometric channels. In other words,
$a^i(\alpha)$ and $b^i(\alpha)$ are the
transmission factor between the photometric flux and the
interferometric and photometric channels, respectively.  Note that
the definition of $a^i(\alpha)$ and $b^i(\alpha)$ includes the transmission of
the beam splitter.}\label{fig_dessin_interf}
\end{figure}
\begin{figure}
    \includegraphics[width=8cm]{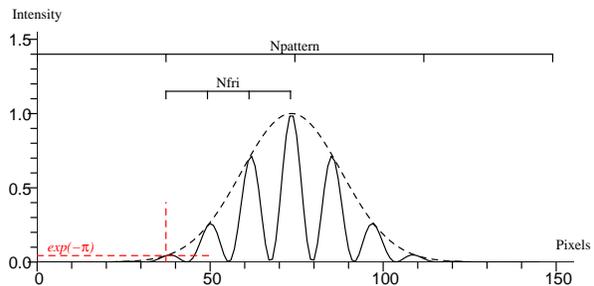}
    \caption{Image of
        a typical interferogram at the output of multiaxial all-in-one
        recombiner. Here is shown a 2-beam recombination with
        instrumental contrast set to $1$. The fringes are weighted by
        the {\bf intensity mode pattern (dashed line). $N_{pattern}$ 
        indicates the width of the detector reading window, the width of one mode pattern being defined as the interval between the top of the Gaussian mode and where the amplitude of the Gaussian mode has decreased by a factor $\exp(-\pi)$. This definition has been chosen so that this width corresponds to half of the first lobe of the Airy pattern in case of pupils which are not weighted by the first mode of single-mode fibers.    
$N_{fri}$ refers to the number
        of fringes in one mode pattern.}This term is
        fixed by the distance between the output pupils.}\label{fig_dessin_fringes}
\end{figure}
Figure \ref{fig_dessin_interf} sketches the principle of multiaxial
recombination in waveguided interferometers. The light arising from the
$i^{th}$ telescope is filtered by a
single mode fiber to convert phase fluctuations of the corrugated
wavefront into intensity fluctuations. The fraction of light $\rho^{i}$
entering the fiber is called the coupling coefficient
\citep{shaklan_1} and depends on the strehl ratio
 {\bf of the apodized pupil} \citep{foresto_1}. Making use of a beam splitter, one part of the
light is selected to estimate the photometry, thanks to dedicated photometric channels. The remaining part of the
light is recombined with the beam coming for the $j^{th}$ telescope to
form fringes. The coding frequency of the fringes $f^{ij}$ is fixed
by the separation of the output pupils, which are arranged along one dimension.

When the only $i^{th}$ beam is lighted,
 the signal recorded on the interferometric channel is the photometric
flux $F^i$ spread on the {\bf intensity mode} pattern $a^i(\alpha)$, that is the
 diffraction pattern of the  $i^{th}$ output pupil weighted by the
 single mode of the fiber. $\alpha$ is the angular variable
 in the image plane. $F^i$ results in 
the source photon flux $N$ attenuated by the total transmission of the
instrument, i.e. the product of the
"static" transmission $t^{i}$ and the coupling coefficient
$\rho^{i}$ of the single mode fiber:
\begin{equation}
F^i = N t^{i} \rho^{i} \label{eq_fi}
\end{equation}

When  beams $i$ and $j$ are lighted simultaneously, the coherent
addition of both beams results in an interferometric component
superimposed to the photometric continuum. The interferometric part,
that is the fringes, arises from the amplitude modulation of the coherent flux
$F_c^{ij}$ at the coding frequency $f^{ij}$. The coherent flux is the
  geometrical product of the photometric fluxes, weighted by the visibility:
\begin{equation}
F_c^{ij} = 2 N \sqrt{t^it^j}\sqrt{\rho^i\rho^j} V^{ij} \mathrm{e}^{i(\Phi^{ij} +
  \phi_p^{ij})} \label{eq_fc}
\end{equation} 
where  $V^{ij}\mathrm{e}^{i\Phi^{ij}}$ is the complex modal 
visibility \citep{mege_1} and $\phi_p^{ij}$ takes into account a
potential differential atmospheric piston. Note that strictly
speaking the modal visibility is not the source visibility. {\bf Rigorously, the modal visibility depends on the convolution between the source visibility and the telescope transfer function which is atmosphere dependent in the optical range. As such the modal visibility is biased both by the geometric antenna-lobe effect (the object is multiplied by the telescope point spread function, as commonly known in radio-astronomy), and by the turbulence. When the object is unresolved by one single telescope however, the modal visibility can be fairly approximated by the object one \citep{tatulli_2} and its estimation is robust (stable to the level of $1\%$ or less) to a change of atmospheric during the calibration process \citep{tatulli_1}. In any case, 
a further study of the relationship between the modal visibility and the source visibility is beyond the scope of this paper, and further informations
can be found in papers mentioned above.} 
Here we consider our observable to be the modal visibility.

Such an analysis can be done for each pair of telescopes available in the
interferometer. 
As a result, the interferogram recorded on the detector can be written
in the general form:
\begin{eqnarray}
&&i(\alpha) = \sum_{i}^{N_{tel}} a^i(\alpha) F^i +\nonumber \\
&&\sum_{i <j}^{N_{tel}} \sqrt{a^i(\alpha)a^j(\alpha)} C^{ij}_{B}(\alpha)\mathrm{Re}\left[F_c^{ij}\mathrm{e}^{i(2\pi\alpha{f^{ij}} + \phi^{ij}_s(\alpha) + \Phi^{ij}_B(\alpha))}\right]\label{eq_interf_general}
\end{eqnarray}
where $\phi^{ij}_s(\alpha)$ is the instrumental phase taking
into account possible misalignment and/or differential phase
between the beams $a^i(\alpha)$ and $a^j(\alpha)$. {\bf  $C^{ij}_{B}$ and $\Phi^{ij}_B(\alpha)$ are respectively the loss of contrast and the phase shift due to diffusion and polarization effects\footnote{assuming a non-polarized incoming light}, which may not be homogeneous along the fringe pattern.}
Fig. \ref{fig_dessin_fringes} gives an
example of a multiaxial all-in-one interferogram in the two telescope
case.

Thanks to the photometric channels, the number of photoevents
$p^i(\alpha)$ coming from each telescope can be estimated independently:
\begin{equation}
p^i(\alpha) =  F^i b^i(\alpha) \label{eq_photo_general}
\end{equation}
where $b^i(\alpha)$ is the detected beam in the $i^{th}$ photometric channel.



We can notice from Eq.'s (\ref{eq_fi}) and (\ref{eq_fc}) that 
the estimator of the squared modal visibility
$\widetilde{|V^{ij}|^2}$  results in the ratio between 
the {\bf squared} coherent flux and the photometric fluxes. 
Using the previous definitions, we can set a
generic form of the estimator as following:
\begin{equation}
\widetilde{|V^{ij}|^2} =
\frac{<|F_c^{ij}|^2>}{<4F^iF^j>} \label{eq_genericvis}
\end{equation}
Note that $<|F_c^{ij}|^2>$ is computed instead of $<F_c^{ij}>$ because
in absence of fringe tracking 
the random atmospheric differential piston $\phi_p^{ij}$ totally blur
the coherent signal. 
It now remains to estimate $F^i$ and $F_c^{ij}$ from the
interferogram. 



\subsection{Fourier estimator: integrating the {\bf power} spectral density} \label{sec_fourier}

In the Fourier space, the interferogram defined by 
Eq. (\ref{eq_interf_general})
takes the form of the sum of photometric and interferometric
peaks. The photometric peaks are centered at the zero spatial frequency
whereas the interferometric peaks {\bf $\widehat{M}_{+}^{ij}(f)$ are located at their
respective spatial coding frequency $f^{ij}$, their counterpart $\widehat{M}_{-}^{ij}(f)$ being in the negative spatial frequency domain.}
{\bf This method in the Fourier space requires that (i) the photometric peaks and the interferometric peaks are not overlapping, and, (ii) the high frequency peaks are not overlapping between each other. If these conditions are fulfilled}, the squared coherent flux can be estimated by
computing the integral of the {\bf power} spectral density $|\widehat{M}_{+}^{ij}(f)|^2$ \citep{roddier_lena_1,conan_1}, {\bf over its frequency support, that is  $[f^{ij} - D/\lambda, f^{ij} + D/\lambda]$ where $D$ is the diameter of the output pupil}.
From the definition of the coherent flux and using the Parseval
equality, it comes:
\begin{equation}
\int |\widehat{M}_{+}^{ij}(f)|^2 \du{f} = |F_c^{ij}|^2  \frac{\int {C^{ij}_{B}}^2(\alpha) a^i(\alpha)a^j(\alpha) \du{\alpha}}{4} \label{eq_contFc}
\end{equation}
The photometric flux is easily computed from the photometric
channel (see Eq. (\ref{eq_photo_general})):
\begin{equation}
\widehat{P}^i =  F^i \int b^i(\alpha) \du{\alpha} \label{eq_contFi}
\end{equation}
Then the estimation of the fringe contrast $C^{ij}$ writes:
\begin{equation}
\widetilde{|C^{ij}|^2} = \frac{<\int |\widehat{M}_{+}^{ij}(f)|^2
  \du{f}>}{<\widehat{P}^i\widehat{P}^j>} = \widetilde{|V^{ij}|^2}.C_{r}^2
\end{equation}
with 
\begin{equation}
C_{r}^2 = \frac{\int  {C^{ij}_{B}}^2(\alpha) a^i(\alpha)a^j(\alpha)\du\alpha }{\int
    b^i(\alpha) \du\alpha \int
    b^j(\alpha) \du\alpha} \label{eq_cr}
\end{equation}
being the instrumental contrast of the recombiner that depends {\bf on the contrast loss due to polarization effects}, on the 
alignment of the beams $a^i(\alpha)$ and $a^j(\alpha)$, and on the
flux ratio between the interferometric and the photometric channels. 
Note that the {\bf power} spectral density of the interferogram has to be properly
unbiased from photon and detector noise \citep{perrin_1}.


\subsection{Model-based estimator: modelling the interferogram}
\label{sec_estim_p2vm}

The model-based estimator has been introduced 
for the first time in the data reduction process of the AMBER
instrument \citep{millour_etal_1}. It consists
in modelling the interferogram thanks to {\it
  a priori} knowledges of the instrument. The purpose of such a signal
processing is twofold: (i) to develop optimized algorithms in terms of
performances of the instrument, that is the SNR of the visibility; and
(ii) on the contrary of the Fourier estimator, to authorize high
frequency peak overlapping when dealing with multi-beam ($N_{tel} >
3$) recombination, thus allowing to code the interferogram on less
pixels. This second point is beyond the scope of this paper. Let
just mention here that plainly choosing the different coding
frequencies is crucial to design optimized multi-beam recombiner making use
of integrated optics. This is especially true in the case of imaging
instruments such as VITRUV \citep{jblebou_etal_2} 
that are recombining $4$ beams or
more, as it is shown in our second paper on the subject
\citep{jblebou_etal_1}.

A full description of this estimator can be found in
\citet{millour_etal_1}. We only recall here the basics principles.
To model the signal on the detector, Eq. (\ref{eq_interf_general}) has
to be rewritten in its sampled version, where $k$ stands for the pixel number, between $1$ and $N_{pix}$:
\begin{equation}
i_k = \sum_{i}^{N_{tel}} F^i a^i_{k} + \sum_{i < j}^{N_{tel}}
c_k^{(i,j)}R^{ij}+d_k^{(i,j)}I^{ij}  \label{eq_ik}
\end{equation}
with
\begin{equation}
c_k^{(i,j)}=\frac{C^{ij}_{B}(k)\sqrt{a^i_{k}a^j_{k}}}{\sqrt{\sum_k {C^{ij}_{B}}^2(k) a_k^{i} a_k^{j}}}\cos(2\pi\alpha_k{f^{ij}}+ \phi^{ij}_s(k) + \Phi^{ij}_B(k))
\end{equation}
and 
\begin{equation}
R^{ij} = \sqrt{\sum_k {C^{ij}_{B}}^2(k) a_k^{i} a_k^{j}} \mathrm{Re}\left[F_c^{ij}\right]
\end{equation}
$d_k^{(i,j)}$ and $I^{ij}$ being the quadratic counterpart of $c_k^{(i,j)}$ and $R^{ij}$ respectively.
$c_k^{(i,j)}$ and $d_k^{(i,j)}$are called the carrying waves of the signal
at the coding frequency $f^{ij}$, since they "carry" (in terms of
amplitude modulation) 
$R^{ij}$ and $I^{ij}$, which are directly linked to the complex coherent
flux.
Furthermore, the photometric fluxes are still computed from the photometric
channels (see Eq. (\ref{eq_contFi})): 
\begin{equation}
p^i = F^i \sum_k b_k^i \label{eq_Fi}
\end{equation}
$F^i$ and $F_c^{ij}$  are then jointly
estimated from the photometry ($p^i$) and the interferogram ($i_k$) by
resolving a set of ($N_{pix}+N_{tel}$) linear equations with ($2N_b +
N_{tel}$) unknowns ($N_b$ being the number of pairs of telescopes, i.e. $N_b=N_{tel}(N_{tel}-1)/2$):
\begin{equation}
\left[ \begin{array}{c} i \\ p \end{array} \right]= [\bf C] . \left[ \begin{array}{c} R\\I\\F \end{array}\right]
\end{equation}
where the matrix ${\bf C}$ takes the detailed form:
\begin{equation}
\overbrace{\left(\begin{array}{ccc}
 \unldots&  c_1^{(i,j)} &\unldots\\ \vert &\deuxvdots&\vert\\
  \unldots  &c_{N_{pix}}^{(i,j)} &\unldots \\ 0 & \unldots & 0 \\
 \vert & \ddots   &\deuxvdots \\0 & \unldots & 0
   \end{array}\right.}^{N_b}\overbrace{\left. \begin{array}{ccc}
 \unldots&  d_1^{(i,j)} &\unldots\\ \vert &\deuxvdots & \vert\\
  \unldots  &d_{N_{pix}}^{(i,j)} &\unldots\\ 0 & \unldots & 0 \\
 \vert & \ddots   &\vert \\0 & \unldots & 0
   \end{array}\right.}^{N_b}\overbrace{\left. \begin{array}{ccc}
 \unldots&  a_1^{i} &\unldots\\ \vert &\deuxvdots&\vert\\
  \unldots  &a_{N_{pix}}^{i} &\unldots\\\sum b_k^1 & \unldots & 0 \\
 \vert & \ddots   &\vert \\0 & \unldots & \sum b^{N_{tel}}_k
   \end{array}\right)}^{N_{tel}}
\end{equation}
The  matrix ${\bf C}$  entirely characterizes the
instrument. It depends on the shape of the detected beams $a^i_k$, $b^i_k$
and on the carrying waves $c_k^{(i,j)}$,  $d_k^{(i,j)}$ that hold
informations about the interferometric beam $\sqrt{a_k^ia_k^j}$, the
coding frequencies $f^{ij}$ and the instrumental differential phases
$\phi^{ij}_s$, and the polarization state within $C^{ij}_{B}$ and $\Phi^{ij}_{B}$. Such quantities can be calibrated in laboratory, hence they are
assumed to be perfectly known. The calibration
procedure is fully described by
\citet{millour_etal_1}. ${\bf C}$ has to be inverted in order to
solve the system. 
In the AMBER experiment, the generalized inverse of ${\bf C}$ has been
called the Pixel To Visibility Matrix (P2VM), since it enables to
compute the visibility of the fringes from the measurements on the detector.
The estimation of the contrast writes: 
\begin{equation}
\widetilde{|C^{ij}|^2} =
\frac{<{R^{ij}}^2> + <{I^{ij}}^2> }{< p^i p^{j}>} = \widetilde{|V^{ij}|^2}.C^2_r
\end{equation}
$C^2_r$ is still the squared instrumental contrast:
\begin{equation}
C^2_r = \frac{\sum_k {C^{ij}_{B}}^2(k) a_k^{i} a_k^{j}}{\sum_k
b^{i}_k\sum_k b^{j}_k} \label{eq_cr_p2vm}
\end{equation}
with the same definition than in Eq. (\ref{eq_cr}).
Note that the quantity ${R^{ij}}^2 + {I^{ij}}^2$ has to be
properly unbiased, like the {\bf power} spectral density in the Fourier plane.

\subsection{SNR of the modal visibility} \label{sec_snr}
Using second order expansion of \citet{papoulis_1}, we
derive from  Eq. (\ref{eq_genericvis})  
the relative error (i.e. the inverse of the SNR) of the squared
visibility:
\begin{eqnarray}
\mathcal{E}^2(|V^{ij}|^2) &=&
\frac{\sigma^2(|F_c^{ij}|^2)}{\overline{|F_c^{ij}|^2}^2} +
\frac{\sigma^2(F^i)}{\overline{F^i}^2} +  \frac{\sigma^2(F^j)}{\overline{F^j}^2} \nonumber \\
&+& 2 \frac{\mathrm{Cov}(F^i,F^j)}{\overline{F^i}\overline{F^j}}-
2\frac{\mathrm{Cov}(|F_c^{ij}|^2, F^iF^j)}{\overline{|F_c^{ij}|^2}\overline{F^i}\overline{F^j}} \label{eq_snr_vis_papoulis}
\end{eqnarray}
The main  difference between both approaches lies in
the following remark: in the case of the Fourier
estimator, the coherent and photometric fluxes are directly estimated 
 from the measurements, each {\it independently}, whereas in the case of the
model-based estimator, the coherent and photometric fluxes are {\it jointly}
  reconstructed from the measurements by way of computation of the
P2VM matrix, and are therefore correlated. As a
result we have the following situations:
\begin{center}
\begin{tabular}{c|c}
Fourier estimator & Model-based estimator \\ \hline \\ 
$\mathrm{Cov}(|F_c^{ij}|^2, F^iF^j) = 0$ &
$\mathrm{Cov}(|F_c^{ij}|^2, F^iF^j) \neq 0$ \\
$\mathrm{Cov}(F^i,F^j) = 0$ & $\mathrm{Cov}(F^i,F^j) \neq 0$ \\  
\end{tabular}
\end{center}
Detailed  computation of Eq. (\ref{eq_snr_vis_papoulis}) are given in
Appendix \ref{app_p2vm} for both estimators, assuming that the 
interferogram is corrupted by photon and detector noise. Atmospheric noise
is neglected here since it has been shown in \citet{tatulli_1} 
that in presence of {\bf modal filtering -- and on the contrary of multispeckle interferometry \citep{goodman_1} -- ,
speckle noise is the dominant noise in the case of very bright sources (negative magnitudes) only, and is therefore marginally relevant. Furthermore, we did not take into account in these computations the effect of atmospheric piston since it only results in the attenuation of the squared visibility, the sensitivity of both estimators to this specific point being the same.}

In next section, we propose to simulate the
estimation of the modal visibility from multiaxial recombination and to
validate our theoretical calculations. Then we compare the
performances of both estimators.

\section{Validation of the theoretical expressions} \label{sec_valid}
\begin{figure*}
    \includegraphics[width=14cm]{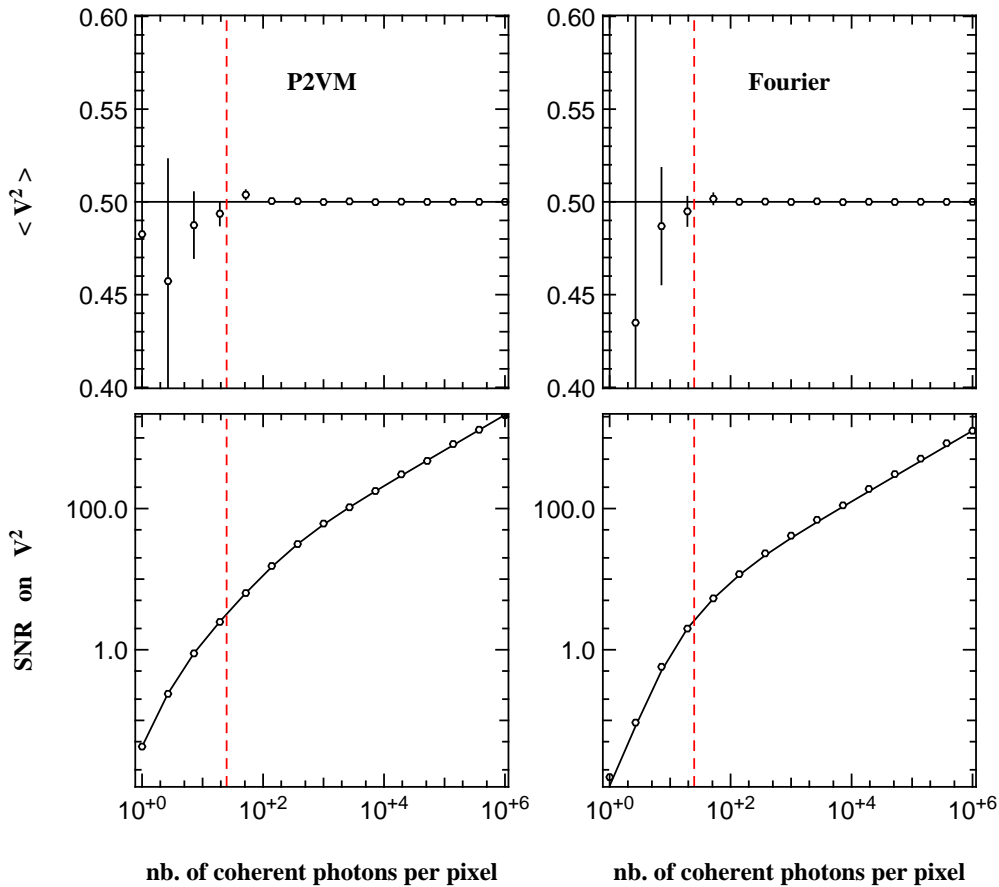}
    \caption{Theoretical (solid lines) and simulated (cross) values of
      the modal visibility mean value (top) and SNR (bottom), for the
      P2VM (left) and the Fourier (right) estimators. Plots are shown
      as a function of the number of photoevents per pixel in the
      interferogram. Detector noise has been set to
      $\sigma=15\mathrm{e}^{-}/\mathrm{pix}$. The vertical dashed line
      shows the limit between the detector and the photon noise
      regime. For the estimation of the visibility arising from
      numerical simulations, we plot the statistical error bars due to
      limited number of samples ($1000$ data sets), that is the
      dispersion of the $1000$ estimated modal visibility, divided by
      the square root of the number of samples.} \label{fig:valid}
\end{figure*}
In order to validate our theoretical expressions derived in the
previous section, we perform statistical simulations of noisy
interferograms. For sake of simplicity we assume from now on a two
telescope interferometer. Note however that 
the validity of our theoretical approach has been
also checked for increasing number of telescopes. Moreover, a deep
analysis of multi-beam ($\ge 3$) recombination is proposed in our second
paper on the subject \citep{jblebou_etal_1}.   

Following the formalism of Section \ref{sec_multi}, the interferogram
arising from multiaxial all-in-one recombination 
is entirely defined by the following parameters: 
\begin{enumerate}
\item The shape of $a(\alpha)$, which is assumed to be the same
  for each beam. It arises from the inverse Fourier Transform of the
  auto-correlation of the output pupil of diameter $D$. {\bf To take into account the weighting of the single mode fiber we assumed a Gaussian shape with a pupil stop of width $D$ (that is a convolution by a Bessel function in the detector plane). 
As a result,  the low and high frequency peaks 
have truncated Gaussian shapes with a base width of $2D/\lambda$.}  We consider
that the photometry is recorded on one pixel.
\item The number of fringes $N_{fri}$ in the interference pattern for
  the lowest coding frequency. It is defined as the distance between
  the closest output pupils in $D/\lambda$ units (see
  Fig. \ref{fig_dessin_fringes}). We choose here $N_{fri}=2$.
\item The number of pixels per fringe $N_{pf}$
  to code the interferogram. It must be
  chosen such that it fulfills the Shannon criteria for the
  highest frequency coding $f_{max}$ of the carrying waves. If it
  is written under the form $f_{max} = \beta D/\lambda$, 
  then the number of pixels must verify: $N_{pf} \ge 2\beta$. For the
  two telescope case considered here, we
  arbitrarily set $N_{pf}=2.5$.
\item The width of the detector reading window $N_{pattern}$ which fixes
  the total number of fringes taken into account in the
  interferogram and hence the total number of pixels $N_{pix}$ read on the
  detector. We impose in this section $N_{pattern}=2$. 
  This choice means that the detector reading window is 
{\bf two mode patterns wide,
exactly as if we only would consider the fringes in the 
first lobe of the diffraction pattern, in the case of an Airy disk.}
 Such choice seems reasonable at first
thought since outside this lobe the interferogram is severely
attenuated (as one can notice in Fig. \ref{fig_dessin_fringes}). Nevertheless, a deeper analysis of this specific point
shows that such a parameter is a key
issue, as it will be discussed 
in Section \ref{sec_def_fov}.   
\item The fraction of flux going into the photometric channels (that
  is the transmission of the beam splitter). We assume here that the
  beam-splitter selects $30\%$ of the flux for the photometry.
\end{enumerate} 
For a given source magnitude, the number of photoevents occurring on
each pixel of the interferometric and the photometric channels are
computed following Eq.'s (\ref{eq_interf_general}) and
 (\ref{eq_photo_general}) respectively, assuming photon noise and
additive detector noise with $\sigma=15\mathrm{e}^{-}/\mathrm{pix}$. Such
a procedure is then repeated until we obtain a  sample of
$1000$ data sets, which is large enough to perform statistics.  
For both estimators, we compute theoretical and statistical mean
value and Signal to Noise Ratio 
of the modal visibility $\widetilde{|V^{ij}|^2}$, 
thanks to Eq.'s (\ref{eq_genericvis}) and (\ref{eq_snr_vis_papoulis}).
Fig.~\ref{fig:valid} shows the results of our computations, for both
methods (formal and simulated) and for both estimators (P2VM and
Fourier), setting the true value of the modal visibility to $0.5$. 
Theoretical calculations and numerical simulations are in
excellent agreement, both for the estimated visibility and the SNR. 
This study validates the theoretical expressions of 
both estimators as well as their respective theoretical SNR. 

\section{Discussion}\label{sec_compare}
\subsection{Estimator relative performances}  
Although both estimators arise from the same formal definition of
Eq. (\ref{eq_genericvis}), they exhibits fundamental conceptual differences.  
Obviously, both techniques presents the same instrumental contrast (see Eq.'s
(\ref{eq_cr}) and (\ref{eq_cr_p2vm}) respectively), which is not
surprising since the instrumental design is strictly the same, as well
as the same estimation of the photometry (see Eq.'s (\ref{eq_contFi}) and (\ref{eq_Fi}) respectively).  However the very
difference lies in the computation of the coherent flux.  

\begin{figure}
     \includegraphics[width=7.5cm]{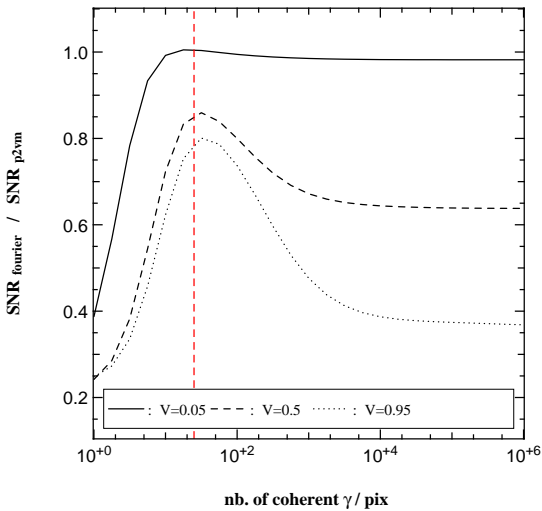}
      \caption{Ratio between the SNR of the Fourier estimator and the
        SNR of the model-based one, 
        as a function of the number of photoevents per pixel in the
      interferogram. Detector noise is still
      $\sigma=15\mathrm{e}^{-}/\mathrm{pix}$. The curves are plotted for
      $3$ types of sources: fully resolved ($V = 0.05$, solid line),
        moderately resolved ($V = 0.5$, dashed line) and unresolved ($V = 0.95$, dotted line).}\label{fig:comp}
\end{figure}

First, in the Fourier case, the coherent flux results in a second order
(i.e. quadratic) estimation, that is $|F_c^{ij}|^2$ is directly
estimated.
On the contrary, the model-based computation is equivalent to a first order
estimation, that is the complex quantity $F_c^{ij}$ is calculated, and then
the squared modulus is taken to get rid of the atmospheric differential
piston. {\bf This latter method is equivalent to fit the complex Fourier Transform of the interferogram. A second order estimator based on modelling would have been the fitting of the power spectral density itself, or equivalently, the fitting of the auto-correlation of the interferogram in the detector plane. Performing first order estimation is above all interesting because it allows to separate informations (i.e. the visibilities for each baseline) before computing the modulus square. As a result, this method prevents ``cross-talk'' between the baselines even if the peaks are partially superimposed, which is particularly worthwhile in the case of visibility estimation from multi-beam ($\ge 3$) interferometers, as developed in our second paper on the subject \citep{jblebou_etal_1}. We also infer that a first oder estimation drive to better -- or at worst identical -- performances than quadratic one, though a thorough analysis of this point, which is beyond the scope of this paper, remains to be done.}
Moreover and above all, the model-based algorithm, thanks to the P2VM
 calibration matrix, takes entire benefit of the knowledge
of the instrument whereas the Fourier estimator does not 
\footnote[1]{although some approximate
  assumptions about the peak position have to be made to set the integration
  interval.}.
When making use of the model-based estimator, 
the shape of the interferogram is perfectly known, precisely its
envelope (the diffraction pattern $\sqrt{a_k^ia_k^j}$) as well as its
coding frequency and its instrumental phase. These {\it a priori}  
informations are gathered in the matrix ${\bf C}$
defined in Section \ref{sec_estim_p2vm}.  
And clearly, introducing
perfectly known (i.e true and un-noisy)  {\it a priori} 
in the data reduction procedures can only improve the performances of
 the corresponding estimator.
These two remarks, and particularly the second point, explain why the
model-based algorithm leads to better performances than the Fourier one, 
as it is illustrated in Figure \ref{fig:comp}. 
Note that the SNR improvement is all the more important than the
visibility is high. In the two telescopes case, a gain of a factor $3$
to $5$ can be achieved at best, for unresolved sources. 
 
However this analysis assumes perfect calibration of the instrument. 
It means that the calibration matrix must be both 
perfectly stable in time and very precise, that is recorded with a SNR
much higher than the SNR of the interferograms.
If the instrument is not stable between the calibration procedures and the
observations, the P2VM will drift and as a result, the estimated
visibilities will be biased. And if the calibration is not precise enough, it
will be the limiting factor of the visibility SNR. In the case of the
AMBER instrument, the calibration procedure is quite complex and it can
require a consequent integration time (several minutes) to get a
useful ad precise calibration. More
generally, the time and the way to calibrate an instrument 
severely depends on its stability and on  
its complexity. Ambitious designs such as the "silicon
v-groove array" of the MIRC recombiner \citep{monnier_2}, 
or recombination schemes making
use of integrated optics chips 
\citep{berger_etal_1, jblebou_etal_2} should drive to
drastic improvements on this specific point.

\subsection{Influence of instrumental parameters}  \label{sec_def_fov}

Previous analysis has been done with a given configuration of the
instrument.  One last point to investigate is how
both estimator behave, one compare to the other, when the parameters
governing the interferogram are varying. Obviously, modifying the 
number of pixel per fringe or even the
detector noise level will result in similar changes for both
estimators, that is the slope of the SNR in the detector noise
regime. 
As well,
changing the coding frequency, which only defines the position of the 
interferometric peak in the Fourier space, will lead to
equivalent modifications of both estimators performances, at least as far as
the high frequency peaks are separable\footnote[2]{when $N_{tel} \ge 3$, see
\citet{jblebou_etal_1}}. 
At last,
choosing the optimized area on which the interferogram gives valuable
informations without adding too much detector noise is
a crucial point. Since the model-based algorithm takes into account the
shape of the interferogram whereas the Fourier estimator does not, the
compromise to find is not the same in both cases. It means the
response to a change of the detector reading window
will differ with regards to the chosen estimator.
The effects of this parameter are investigated here.

\begin{figure*}
  \begin{center}
    \includegraphics[width=13.5cm]{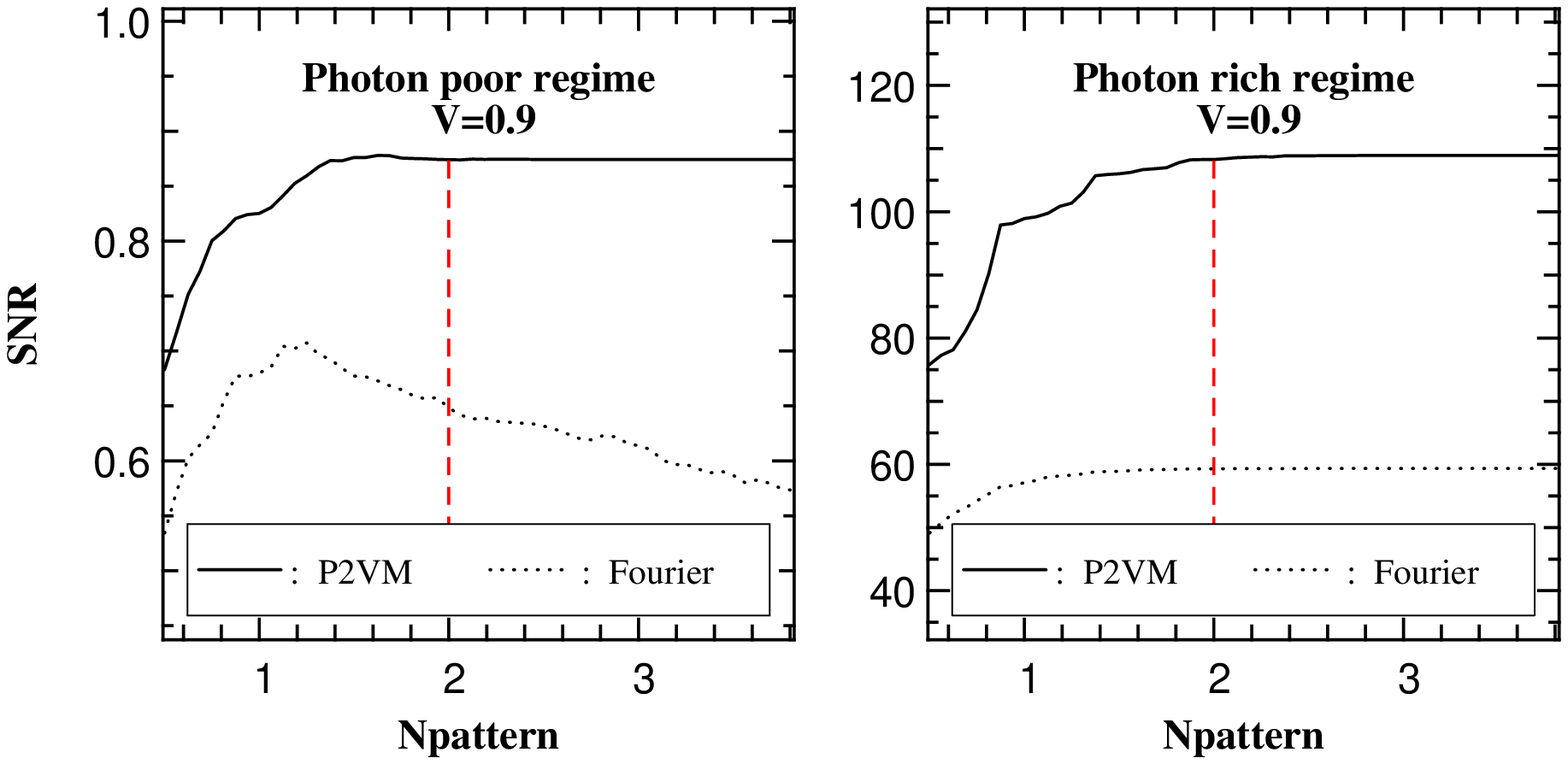}
    \includegraphics[width=13.5cm]{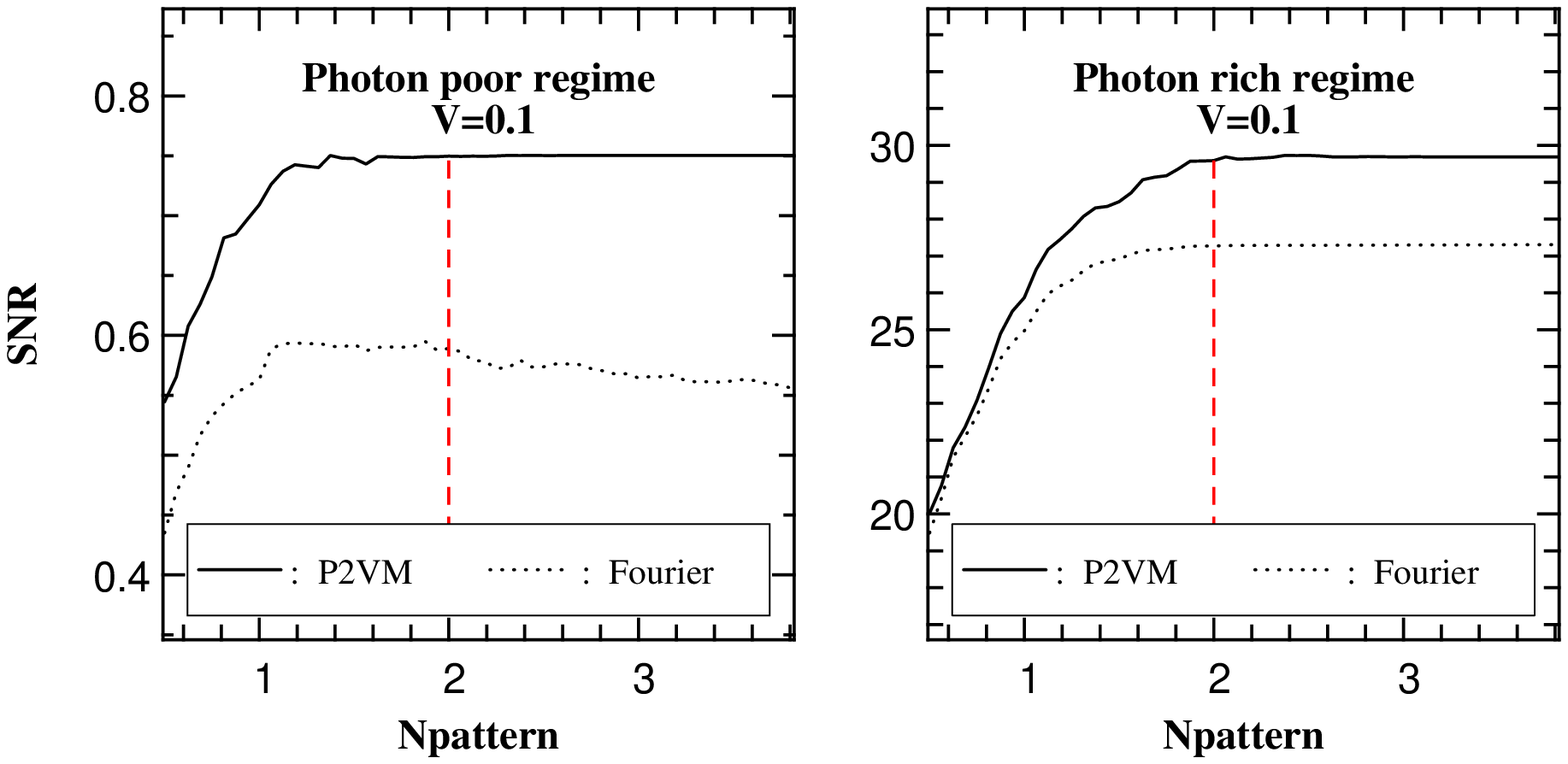}
    \caption{SNR of the modal visibility as a function 
      of the detector reading window, in fraction of {\bf intensity mode} pattern, 
      for photon poor (left, here $1\gamma/\mathrm{pix}$ ) and photon rich (right, here $10^{6}\gamma/\mathrm{pix}$) regimes. Results are
      displayed for marginally resolved 
      ($V=0.9$, top)  and fully resolved ($V=0.1$, bottom) sources.
      The instrumental parameters 
      are: $N_{fri} = 4$ fringes per beam, 
      $N_{pf} = 4$ pixels per fringe, and a detector noise of $\sigma=15\mathrm{e}^{-}/\mathrm{pix}$. The vertical dash
lines correspond to recording one entire mode pattern (see Fig.~\ref{fig_dessin_fringes}).
\label{fig:fov}}
  \end{center}
\end{figure*}

In a multiaxial combination, one has the choice of the limits of the
reading window on the detector, i.e of the number of pixels to
consider. 
And the largest the window, the more signal you
integrate, but the more detector noise you record too. 
In Figure \ref{fig:fov}, we show the evolution of the SNR ratio of the
modal visibility as a function of the width of the window on the
detector (here defined in fraction of {\bf intensity mode} pattern). 
The entire first lobe of the fringe pattern contains $N_{fri} = 8$
fringes, with $N_{pf} = 4$ pixels per fringe.
All the other parameters of the instrument are kept unchanged.


{\it Fourier estimator:} In the photon poor regime, at the detection
limit of the instrument, the SNR shows a maximum when the width of the
reading window is about half of the first lobe of the envelope.
Beyond that point, the pixels 
have very small individual SNR and only bring
a noise contribution in the estimation of the visibility.
In the photon rich  regime, the SNR reaches its maximum 
for the same width of the reading window but exhibits a plateau
as the width increases. This behavior stands 
as long as the SNR of each individual
pixel is dominated by the photon noise. Then the SNR of the visibility
starts to decrease. 
The width of the plateau depends on the incoming flux of
the source and is all the more large than the source is bright.

{\it Model-based estimator:} As for the Fourier estimator, the shape of the SNR is linked to the
shape of the fringe pattern. But in this case, the SNR is increasing
with the width of the detector window and does not exhibit a maximum
(in other word, the optimal is found for an "infinite"
width). 
As a matter of fact, thanks to the generalized inverse of the matrix
${\bf C}$ that takes into account the shape of the interferogram, each pixel
contribution is weighted by its individual SNR. 
So the pixels with bad SNR (due to the envelope or fringe
modulation) are "removed" from the reconstruction and do not introduce
noise in the estimation of the visibility. 
Nevertheless, we can see that the slope of the SNR 
becomes almost flat from a detector width of about one {\bf intensity mode} pattern.

This analysis shows that, 
in the framework of interferometric observations making use of
multiaxial all-in-one recombination, and in the case
of bright sources, it is worth the effort to integrate the  {\bf
interferogram on the entire lobe} in order to
optimize the SNR of the visibility. 
This statement stands for both estimators.   
When observing faint sources, that is when reaching the limiting
magnitude of the instrument, and in the specific case of the Fourier
estimator, performances are slightly improved when reducing the width of the
reading window to half of the lobe, although the gain on the SNR
never exceed a factor of $2$.



\section{Conclusion} \label{sec:discussion}
In this paper, we have developed the theoretical formalism that allows
to model single mode interferometers using multiaxial
all-in-one-coding, from the signal processing point of view. 
From this formalism, two estimators of the
visibility have been analyzed. The first one consist in using the classical
integration of the {\bf power} spectral density of the interferogram in the Fourier
plane whereas the second one deals with modeling the interferogram in
the detector plane, as it has been chosen for the AMBER experiment. 
Performances of such estimators  
have been computed. Considering photon and detector noises, 
theoretical expression of the SNR of the visibility have been
recalled for the Fourier estimator, and derived for the first time 
in the case of model-based estimators.
These expressions have been validated 
through numerical simulations and then compared. 
We have shown that the second technique 
offers optimal performances since it makes full use of
knowledges about the instrument, especially the shape of the
interferogram. In the two telescopes case which has been
emphasized in this paper, we have demonstrated that the model-based estimator
enables at best to come over a factor of $5$ of the visibility SNR,
compare to the Fourier one. Finally we have addressed the
question of the width of the reading window of the detector. This
point is indeed a crucial issue when dealing with multiaxial recombination. 
We have shown that, regardless of the
chosen estimator, integrating  the entire lobe of {the intensity mode
pattern offers optimized performances.}


\section*{Acknowledgments}
Authors thank Drs K. Perrault and F. Malbet for very helpful
suggestions that improved the presentation of the paper. They wish also to thank the referee for its careful reading of this work and its subsequent comments that made this paper much clearer. All the calculations and
graphics were performed with the freeware
\texttt{Yorick}\footnote{\texttt{ftp://ftp-icf.llnl.gov/pub/Yorick/doc/index
.html}}

\begin{appendix}

\input{MF479rv_app.tex}
\end{appendix}

\label{lastpage}

\end{document}

%% file: MF479rv_app.tex

\section{Theoretical SNR of the visibility} \label{app_p2vm}
We assume that the interferogram as well as the photometric outputs are
corrupted by photon (Poisson) noise and additive Gaussian noise of
variance $\sigma^2$. 
\subsection{Generic expression}
The estimator of the square visibility can be express in a generic form:
\begin{equation}
\widetilde{|V^{ij}|^2} \propto
\frac{<|F_c^{ij}|^2>}{<F^iF^j>}
\end{equation}
where $F_c^{ij}$  the coherent flux at the frequency $f^{ij}$, an
$F^i$, $F^j$ are the photometric fluxes.

The relative error $\mathcal{E}(|V^{ij}|^2)$ on the square
visibility is then given by \citet{papoulis_1}:
\begin{eqnarray}
\mathcal{E}^2(|V^{ij}|^2) &=&
\frac{\sigma^2(|F_c^{ij}|^2)}{\overline{|F_c^{ij}|^2}^2} +
\frac{\sigma^2(F^i)}{\overline{F^i}^2} +  \frac{\sigma^2(F^j)}{\overline{F^j}^2} \nonumber \\
&+& 2 \frac{\mathrm{Cov}(F^i,F^j)}{\overline{F^i}\overline{F^j}}-
2\frac{\mathrm{Cov}(|F_c^{ij}|^2, F^iF^j)}{\overline{|F_c^{ij}|^2}\overline{F^i}\overline{F^j}} \label{eq_app_papou}
\end{eqnarray}
Theoretical expression of each term of previous equation is now given
for both estimators. 
\subsection{SNR for the Fourier estimator}
The coherent flux is linked to the spectral density of the
interferogram by the following relationship (see Eq. (\ref{eq_contFc}), written
in its sampled form):
\begin{equation}
|F_c^{ij}|^2 \propto \sum_k |\widehat{M}^{ij}(f_k)|^2
\end{equation}
Hence the expected value and the error on the coherent flux writes:
\begin{eqnarray}
\overline{|F_c^{ij}|^2} &\propto&  \sum_k \overline{|\widehat{M}^{ij}(f_k)|^2} \\
\sigma^2(|F_c^{ij}|^2) &\propto& \sum_k \sigma^2(|\widehat{M}^{ij}(f_k)|^2)
+ \nonumber \\
&&\sum_k \sum_{l \ne k} \mathrm{Cov}(|\widehat{M}^{ij}(f_k)|^2,
\widehat{M}^{ij}(f_l)|^2)
\end{eqnarray}
The statistics of the spectral density  of an interferogram  have already
been  computed by \citet{goodman_1} in the case of photon
noise and completed by \citet{tatulli_1} with detector
and atmospheric noise. We
recall the results here, without taking into account the atmospheric
noise (i.e. speckle noise):
\begin{equation}
\overline{|\widehat{M}^{ij}(f_k)|^2} = \overline{N}^2|\widehat{i}(f_{k})|^2 +
\overline{N} + N_{pix}\sigma_{det}^2  \label{eq_meanvis}
\end{equation}
{\bf where we recognize the bias part due to photon noise ($\overline{N}$, \citet{goodman_1}) and additive Gaussian noise ($N_{pix}\sigma_{det}^2$, \citet{tatulli_1}),}
\begin{eqnarray}
\sigma^2(|\widehat{M}^{ij}(f_k)|^2) &=& 2\overline{N}^3|\widehat{i}(f_{k})|^2
+ 4\overline{N}^2|\widehat{i}(f_{k})|^2 + \overline{N}^2 +\nonumber \\
&& N_{pix}^2\sigma^4 + 3N_{pix}\sigma^4 +  2N_{pix}\sigma^2\overline{N} +\nonumber \\
&& 2N_{pix}\sigma^2\overline{N}^2|\widehat{i}(f_{k})|^2
\end{eqnarray}
\begin{eqnarray}
&& \mathrm{Cov}(|\widehat{M}^{ij}(f_k)|^2, \widehat{M}^{ij}(f_l)|^2) =
2\overline{N}^3\mathrm{Re}\left[\widehat{i}(f_k)\widehat{i}^{\ast}(f_l)\widehat{i}^{\ast}(f_k-f_l)\right]
+ \nonumber\\
&&
2\overline{N}^3\mathrm{Re}\left[\widehat{i}(f_k)\widehat{i}(f_l)\widehat{i}^{\ast}(f_k+f_l)\right]
+ 2\overline{N}^2|\widehat{i}(f_k)|^2 + 
2\overline{N}^2|\widehat{i}(f_l)|^2 +  \nonumber\\
&&\overline{N}^2|\widehat{i}(f_k-f_l)|^2 +
\overline{N}^2|\widehat{i}(f_k+f_l)|^2 + \overline{N} + 3N_{pix}\sigma^4
\end{eqnarray}
where $\widehat{i}(f)$ is the normalized spectral density
(such as $\widehat{i}(0) =1$), {\bf that is:
\begin{eqnarray}
\widehat{i}(f) &=& V(f) \widehat{g}(f) \\
\widehat{i}(f) &=& \frac{V(f)}{N_{tel}} \widehat{g}(f) 
\end{eqnarray}
for the photometric and the interferometric peaks respectively, $N_{tel}$ being the number of telescopes, and $\widehat{g}(f)$ being the normalized Fourier Transform of the beam $a(\alpha)$ (assuming, the shame shape for the whole beams).}
Furthermore, we have
\begin{equation}
\sigma^2(F^i) = \overline{F^i} + \sigma_{det}^2
\end{equation}
Finally, since the coherent flux and each photometric flux are
estimated independently, we have:
\begin{equation}
\mathrm{Cov}(|F_c^{ij}|^2, F^iF^j) = 0
\end{equation}
\begin{equation}
\mathrm{Cov}(F^i, F^j) = 0
\end{equation}
\subsection{SNR for the P2VM estimator}
We recall that the real and imaginary part of the weighted complex
visibility are defined by the system of equations:
\begin{equation}
\left[ \begin{array}{c} i \\ P \end{array} \right]= [C] . \left[ \begin{array}{c} R\\I\\F \end{array}\right]
\end{equation}
if we call ${\bf M} = m_k, k \in [1..N_{pix} + N_{tel}]$ the vector resulting in the concatenation of the
interferogram $i$ and the photometry $P$, we can write: 
\begin{equation}
R^{ij} = \sum_k^{N_{pix}} \xi_k^{ij} m_k
\end{equation}
\begin{equation}
I^{ij} = \sum_k^{N_{pix}} \zeta_k^{ij} m_k
\end{equation}
\begin{equation}
F^{i} = \sum_k^{N_{pix}} \beta_k^{i} m_k
\end{equation}
where $ \xi_k^{ij}$,$\zeta_k^{ij}$ and $ \beta_k^{i}$ are the
coefficients of the P2VM matrix.
Hence it comes:
\begin{equation}
|F_c^{ij}|^2 = {R^{ij}}^2 + {I^{ij}}^2 =  \sum_k\sum_l [\xi_k^{ij}\xi_l^{ij}+\zeta_k^{ij}\zeta_l^{ij}]m_km_l
\end{equation}
\begin{equation}
F^iF^j =  \sum_k\sum_l \beta_k^{i}\beta_l^{j}m_km_l
\end{equation}
Here, the covariance between the coherent flux and the photometric
fluxes, as well as the covariance between the photometric fluxes have to
be taken into account. For sake of simplicity, Eq. (\ref{eq_app_papou}) can be
rewritten:
\begin{equation}
\mathcal{E}^2(|V^{ij}|^2) =
\frac{\sigma^2(|F_c^{ij}|^2)}{\overline{|F_c^{ij}|^2}^2} +
\frac{\sigma^2(F^iF^j)}{\overline{F^iF^j}^2} -
2\frac{\mathrm{Cov}(|F_c^{ij}|^2, F^iF^j)}{\overline{|F_c^{ij}|^2}\overline{F^i}\overline{F^j}}
\end{equation}
It now remains to compute all the terms knowing that:
\begin{eqnarray}
\sigma^2(|F_c^{ij}|^2) &=& \overline{|F_c^{ij}|^4} -
\overline{|F_c^{ij}|^2}^2 \\
\sigma^2(F^iF^j) &=& \overline{{F^i}^2{F^j}^2} - \overline{F^iF^j}^2\\
\mathrm{Cov}(|F_c^{ij}|^2, F^iF^j) &=& \overline{|F_c^{ij}|^2F^iF^j}-\overline{|F_c^{ij}|^2}\overline{F^iF^j}
\end{eqnarray}
\begin{table}
\caption{Coefficients of the fourth
    order statistics of the coherent and photometric fluxes.}\label{table_coef}
\begin{equation} 
\begin{array}{|c|c|c|} \hline&&\\
&\overline{|F_c^{ij}|^4} & \overline{{F^i}^2{F^j}^2}  \\&&\\ \hline &&\\
\alpha_{k,l,n,o}^{(1)} & \gamma_{kl}^{ij}\gamma_{no}^{ij} &
\beta^i_k\beta^j_l\beta^i_n\beta^j_o \\ & &\\
\alpha_{kln}^{(2,1,1)}&2\gamma_{kk}^{ij}\gamma_{ln}^{ij} +
4\gamma_{kl}^{ij}\gamma_{kn}^{ij} &
\beta^{i^2}_k\beta^j_l+\beta^{j^2}_k\beta^i_l\beta^i_n + 4
\beta^{i}_k\beta^j_k \beta^i_l \beta^j_n \\ &&\\
\alpha_{kl}^{(2,2)}&
\gamma_{kk}^{ij}\gamma_{ll}^{ij}+2\gamma_{kl}^{ij}\gamma_{kl}^{ij}&\beta^{i^2}_k\beta^{j^2}_l
+ \beta^{i}_k\beta^{i}_l\beta^{j}_k\beta^{j}_l \\ &&\\
\alpha_{kl}^{(3)}&4\gamma_{kk}^{ij} \gamma_{kl}^{ij}&
2\beta^{i^2}_k\beta^{j}_k\beta^{j}_l +
2\beta^{j^2}_k\beta^{i}_k\beta^{i}_l \\ &&\\
\alpha_{k}^{(4)} & \gamma_{kk}^{ij}\gamma_{kk}^{ij} &
\beta^{i^2}_k\beta^{j^2}_k \\ &&\\\hline
 \end{array} \nonumber
\end{equation}\\
\begin{equation}
\begin{array}{|c|c|} \hline&\\
& \overline{|F_c^{ij}|^2F^iF^j} \\&\\ \hline &\\
\alpha_{k,l,n,o}^{(1)}  & \gamma_{kl}^{ij} \beta^i_n \beta^j_o\\ & \\
\alpha_{kln}^{(2,1,1)} &
\gamma_{kk}^{ij}\beta^i_l\beta^j_n +
\gamma_{ln}^{ij}\beta^i_k\beta^j_k + 2\gamma_{kl}^{ij}[\beta^i_k\beta^j_n+\beta^i_n\beta^j_k]\\ & \\
\alpha_{kl}^{(2,2)} &
\gamma_{kk}^{ij}\beta^i_l\beta^j_l + \gamma_{kl}^{ij}[\beta^i_k\beta^j_l+\beta^i_l\beta^j_k]\\ &\\
\alpha_{kl}^{(3)} &
\gamma_{kk}^{ij}[\beta^i_k\beta^j_l+\beta^i_l\beta^j_k] + 2\gamma_{kl}^{ij}\beta^i_k\beta^j_k \\ &\\
\alpha_{k}^{(4)} &\gamma_{kk}^{ij}\beta^i_k\beta^j_k \\ &\\\hline
 \end{array} \nonumber
\end{equation}
\end{table}
To lighten the calculations we introduce the variable $\gamma$ such
that $\gamma_{kl}^{ij}=\xi_k^{ij}\xi_l^{ij}+\zeta_k^{ij}\zeta_l^{ij}$.
Then we can compute the second order statistics of the square coherent
flux and the photometric fluxes:
\begin{equation}
\overline{|F_c^{ij}|^2} =   \sum_k\sum_l
\gamma_{kl}^{ij} \overline{m_km_l} 
 = \sum_k \gamma_{kk}^{ij} \overline{m_k^2} +
  \sum_k\sum_{l \neq k} \gamma_{kl}^{ij} \overline{m_k}~\overline{m_l}
\end{equation}
\begin{equation}
\overline{F^iF^j} =  \sum_k \beta_k^{i}\beta_k^{j}\overline{m_k^2} +   \sum_k\sum_{l \neq k}  \beta_k^{i}\beta_l^{j} \overline{m_k}~\overline{m_l}
\end{equation}
Then the fourth order statistics $\overline{F^4}=\overline{|F_c^{ij}|^4}$, $\overline{{F^i}^2{F^j}^2}$, $\overline{|F_c^{ij}|^2F^iF^j}$ can be described by the generic equation:
\begin{eqnarray}
\overline{F^4}&=& \sum_k \alpha^{(4)}_{k} \overline{m_k^4} +
\sum_k\sum_{l \neq k} \alpha^{(3)}_{kl} \overline{m_k^3}~\overline{m_l}+
\nonumber \\
&& \sum_k\sum_{l \neq k} \alpha^{(2,2)}_{kl} \overline{m_k^2}~\overline{m_l^2}
+ \sum\sum_{l \neq k \neq n}\sum \alpha^{(2,1,1)}_{kln} \overline{m_k^2}~\overline{m_l}~\overline{m_n} +
\nonumber \\
&&\sum\sum_{l \neq k} \sum_{ \neq o \neq n}\sum   \alpha^{(1)}_{klno} \overline{m_k}~\overline{m_l}~\overline{m_n}~\overline{m_o}
\end{eqnarray}
where $\alpha^{(1)}_{klno}$, $\alpha^{(2,1,1)}_{kln}$,
$\alpha^{(2,2)}_{kl}$, $ \alpha^{(3)}_{kl}$, and $\alpha^{(4)}_{k}$
are given in table \ref{table_coef} in each specific case.
{\bf Previous equations can be computed knowing the statistics of $m_k$, that is:
\begin{eqnarray}
\overline{m_k^2} &=& \overline{m_k}^2 + \overline{m_k} + \sigma^2 \\
\overline{p^2} &=& \overline{p}^2 + \overline{p} + N_{photpix}\sigma^2 
\end{eqnarray} 
$N_{photpix}$ being the number of pixels to code the photometric outputs. The third and fourth moments $\overline{m_k^3}$ and $\overline{m_k^4}$ are derived from first and second order ones assuming Gaussian statistics for sake of simplicity.}